\documentstyle[10pt]{article}

\begin{document}

\begin{center}
{\bf  MASS SPECTRA OF RADIALLY AND ORBITALLY
EXCITED STATES OF MESONS}

\bigskip

{ V.Khruschov \footnote{Talk given at 12th Lomonosov Conference on Elementary
Particle Physics, Moscow, August 25-31, 2005 \\
\hspace*{4mm} e-mail: khru@imp.kiae.ru},
V.Savrin \footnote{e-mail: savrin@theory.sinp.msu.ru},
S.Semenov \footnote{e-mail: semenov@imp.kiae.ru}
}

{${}^{a,c}$RRC "Kurchatov Institute", Kurchatov Sq., 
123182 Moscow, Russia \\
${}^b$Skobeltsyn Institute of Nuclear Physics, Moscow State University
}
\end{center}

\begin{abstract}{ Meson mass spectra, evaluated in the framework of the relativistic 
model of quasi-independent quarks, are presented. Mass values are obtained
with the help of numerical calculations based on the Dirac equation and by
phenomenological mass formulae. The Dirac equation involves the potential,
which is sum of the vector quasi-Coulombic potential and the scalar linear
rising confinement potential. 
 The phenomenological mass formulae are applied to excited
meson states consisting of $u-$, $d- $quarks and antiquarks with isotopical
spin $I=1$. A comparison  of the evaluated mass spectra with
 existing data is performed.
Problems of identification of some meson states in vector and scalar
channels are discussed.
}
\end{abstract}

\noindent It is  known that the calculation of hadron mass spectra on the 
level  of experimental data precision  \cite{eidel}  still remains among the 
unsolved QCD problems due to some technical and conceptual difficulties, which
are related mainly with the nonperturbative effects, such as confinement and
spontaneous chiral symmetry breaking. Nowadays 
 hadron characteristics calculations are frequently
carried out with the help of  phenomenological  models.
Among phenomenological  models the potential quark models are the
simplest ones, they  make it possible to represent the different stages of
calculations in  terms of commonly used  physical quantities.  Although these 
models are suitable rather well for the heavy quark
hadrons treating of hadrons containing the light quarks is more
complicated task, which demands  relativistic  and 
nonpotential effects to be taken into account \cite{byk,godf,diek}.
The relativistic model
of quasi-independent quarks is applicable to the description of  
 properties of light and heavy hadrons \cite{bog,bogo,khru,khrusas}.
Below we present the results of  calculations 
in the framework of this model \cite{khrusas,ks,kh} of mass spectra of
quarkonium type mesons, which are orbitally and radially excited states.
We compare results obtained with  data, make predictions for mass
values of unobserved yet meson states
and discuss  possible  structures of some mesons in scalar  and vector  
channels.

According to the main statement of the independent-quark model a hadron is
considered as a system composed of a few non-interacting with each other
directly valency constituents (quarks, 
diquarks and constituent gluons) with the coordinates $\mathbf{r}_{i}$, 
$i=1,...,N$, which moving in some mean 
colour singlet confining field. It is assumed
that the  field is spherically symmetric and its motion in space is
determined by motion of its center with the coordinate $\mathbf{r}_{0}$. 
We treat the mean field as a quasi-classical object possessing
some energy $\epsilon _{0}$. Meantime each of $N$ constituents interacting
with the  mean field gets the state with a definite value of its
energy $\epsilon _{i}$, which can be determinated after solving
of the model equation. We choose  the Dirac equation as the
model equation for  quarks (the Klein-Gordon equation for diquarks and 
constituent gluons):
\begin{equation}
\sqrt{\lambda _{i}+m_{i}^{2}}\psi _{i}(\mathbf{r_{i}})=\left[ 
({\mbox{\boldmath{$\alpha_i$}}}{\mathbf{p_{i}}})+
\beta_{i}({\it m_{i}}+{\it V_{0}})+{\it V_{1}}\right] \psi _{i}(%
\mathbf{r_{i}}),
\label{diq}
\end{equation}
with $E_{i}(n_{i}^{r},j_{i})=$ $\sqrt{\lambda _{i}+m_{i}^{2}}$,
 $i=1,2$,  $V_{0}(r)=$ $\sigma r/2$ and $V_{1}(r)=$ $-2\alpha _{s}/3r$, 
where the model parameters $\sigma $ and $\alpha _{s}$ have meanings of 
the string tension and the strong coupling constant at small distances, 
correspondingly. In order to exclude superfluous meson states, the 
following selection rules for $n{}^{2S+1}L_{J}$ -states
with quark masses $m_{1}$, $m_{2}$ and quantum numbers 
$n_{1}^{r}$, $j_{1}$, $n_{2}^{r}$, $j_{2}$ are used 
\begin{equation}
\begin{array}{l}
n_{1}^{r}=n_{2}^{r}=n-1, \quad j_{1}=j_{2}=J+1/2,\quad if\quad J=L+S, \\ 
j_{1}=j_{2}+1=J+3/2,\quad if\quad J\neq L+S,\quad m_{1}\leq m_{2,}%
\end{array}%
\label{jnf}
\end{equation}

After the elimination the angular dependence the  radial equation obtained
 can be solved only by numerical methods. For
calculating its eigenvalues the  code algorithm, which was based on
the Numerov three point recurrent relation  \cite{num}, has been used 
jointly with the regularization procedure for the singular potential at $r=0$.
In order to estimate the spin-spin interaction between a quark and an antiquark
we use the  expression for this interaction from 
 the relativized quark model  \cite{godf}:
$
V_{S_1S_2}=\frac{32\pi\alpha_s \mbox{\boldmath $s_1$}\mbox{\boldmath $s_2$} 
}{9E_1E_2} \delta^3(\mathbf{r}), 
$
\noindent where $\mbox{\boldmath $s_1$}$ and $\mbox{\boldmath%
$s_2$}$ are the spin operators and $E_1$, $E_2$ are the energies of quark
and antiquark, respectively. Evaluation of the expectation value of this
operator in the first order of perturbation theory 
shows that a contribution of the spin-spin interaction is a slowly varying
function versus $E_1$ and $E_2$ and depends mainly on the total spin of
quark-antiquark system and the $\alpha_s$. For S-wave
mesons we use the  mass formula  \cite{ks}:
\begin{equation}
M_{m}=E_{0m}+E_{1}(n_{1}^{r},1/2)+E_{2}(n_{2}^{r},1/2)+4<%
\mbox{\boldmath
$S_1$}\mbox{\boldmath $S_2$}>_{q_{1}\bar{q}_{2}}V_{SS},
\label{fueq}
\end{equation}
\noindent where $V_{SS}$ is approximately proportional to $\alpha _{s}$ 
with  constant $v_{0S}$. As calculations show, the $E_{0m}$  are negligible
for the majority of $0^{-+}$ and $1^{--}$ mesons, excluding 
 only a few cases, which will be discussed below. It should be noted, that
for ground states of $\overline{q}q^{\prime }-$ mesons the form of relation
(\ref{fueq}) is analogous (after the substitution $E_i\to m_i^c$, $i=1,2$) 
to the mass formula, which depends on constituent quark masses $m_i^c$  
 \cite{zeldsa}.
Although  for vector heavy mesons the contributions of spin-spin interaction 
are on the level of the model absolute uncertainty $\sim 40 MeV$,
nevertheless, for pseudoscalar mesons these contributions should be 
taken into account. The magnitude of the spin-spin interaction constant 
$v_{0S}$ is approximately equal to $100 MeV$. In the case of isoscalar mesons, 
which are neutral with respect to flavour quantum numbers, the  terms due to
the annihilation interaction for quark and antiquark or instanton interaction
appear  \cite{fili, doroho}. But in the framework of our model  
the contributions of such type can be included into $E_{0m}$ terms.

We have found that a suitable fit within the model accuracy for
masses of ground state mesons, which are composed of quark and antiquark
with $u-$, $d-$, $s-$, $c-$ or $b-$ flavours \cite{eidel}, is provided by the
following values of $E_{0m}$ parameter:
$
E_0(\eta _c)= E_(J/\Psi)=-150 MeV, \quad E_0(K)=-200 MeV,
$
$
E_0(\pi ) = E_0(\eta_b) = E_0(\Upsilon) =-450 MeV.
$
For the  $\alpha _s$ 
and the quark masses we obtain the following values:
${\bar{m}}
=(0.007\pm 0.005)$ GeV, 
$\alpha _{s}^{n
}=0.60\pm 0.15,$  
$m_{s}=(0.14\pm 0.03)$ GeV, 
 $\alpha _{s}^{s}=0.47\pm 0.10,$  
$m_{c}=(1.28\pm 0.05)$ GeV, 
 $\alpha _{s}^{c}=0.33\pm 0.03,$ 
$m_{b}=(4.60\pm 0.10)$ GeV, 
$\alpha _{s}^{b}=0.27\pm 0.02$,
$\sigma =(0.20\pm 0.01)GeV^2$.
Within the model accuracy data for the well known experimental mass values
for $1^{--}$ and  $0^{-+}$ mesons with  $u$-, $d$-,
 $s$-, $c$- or $b$ quark and antiquark 
are consistent with  the presented above 
model parameters. Thus in the framework of the model the
flavor independence of the confinement
potential \cite{doro,mart,quroth}
is confirmed as well as the concordance with the asymptotic 
freedom behaviour of the $\alpha _{s}$. The evaluated mass spectra of  mesons 
are displayed in the Table 1.

In spite of  difficulties which exist in for evaluations of mass spectra
of light hadrons  a number of relations among their
masses have been obtained using  phenomenological considerations,
such as the mass relations  for Regge trajectories, 
daughter Regge trajectories \cite{r3,r4,r5,k2} and the mass formulae, 
obtained on the base of finite energy QCD sum rules \cite{k1}. 
In the framework of relativistic quasi-independent quarks model the quark and
antiquark energies are determined from the solutions of 
the Dirac equation. However, for the superlight 
$\overline{q}q^{\prime }-$mesons with 
 $u-$ and $d-$ quarks and antiquarks only, the phenomenological energy 
spectral functions $E_{i}(n_{i},j_{i},c,\kappa )$  are 
suitable within the  relative uncertainty of the order of a few percents
\cite{khru}:
\begin{equation}
M(n^{2S+1}L_{J})=E_{1}(n_{1}^{r},j_{1},c,\kappa)+
E_{2}(n_{2}^{r},j_{2},c,\kappa),  
\label{mnf}
\end{equation}
where 
$$
E_{i}(n_{i},j_{i},c,\kappa)=\left\{ 
\begin{array}{c}
c+\kappa \sqrt{2n^{r}+L+j-1/2},\quad L+j-1/2=2k,\quad k=0,1,... \\ 
\kappa \sqrt{2n_{r}+L+j-1/2},\quad L+j-1/2=2k+1,\quad k=0,1,...%
\end{array}
\right.  
$$

\noindent Note that nonpotential corrections, 
which are connected mainly with $c$ value, are the most important for
mass spectra of light mesons \cite{fili,diek}.  When evaluating masses of 
unobserved yet  excited meson states  we use the formulae
written above together with the values of two parameters $c$ and 
$\kappa $, which have been obtained by fitting the mass values of 
experimentally detected meson states. The $c$ and $\kappa $ values 
obtained by this manner are $c=69$ $MeV,$ $\kappa =382\pm 4$ $MeV$ 
\cite{kh}. The results obtained with $c=69$ $MeV,$ $\kappa =385$ $MeV$ 
are shown in the Table 2, where  the mass values of orbital 
excitations of superlight meson states up to $L=4$ are presented.

The mass formulae presented above
permit to explain the degeneracy of mass values for different $%
J^{PC}$ mesons. For example, we get $M(\rho^{\prime%
\prime})=M(\rho_3)$. If one substitute the data, then $1720\pm20 MeV=1688.8\pm2.1
MeV$. In the high mass value region the degeneracy of such kind will be
increased and it is seen in the $\sim 2315 MeV$ region.
As it follows from the Table 2, the mass value calculated for 
orbital excitation of the  $1^{--}$ meson is equal to $1506 MeV$. 
So  mixing between the 
radial and orbital excitations for the $1^{--}$ mesons should take place,
which leads to the experimentally visible vector meson states 
$V_{R}^{\prime }$ and $V_{O}^{\prime }$.
Taking into account the  experimental data the following cases are
most preferable. In the first case the  observed bump at $%
1465 MeV$ consists of one resonance $V_{R}^{\prime }$, that is mainly the
radial excitation of $\rho -$meson. In the second case the 
observed bump at $1465 MeV$ consist of two resonances $V_{R}^{\prime }$ and 
$V_{O}^{\prime }$ (see also \cite{acko}). 
Another importain problem is the interpretation of mesons discovered in
scalar channel. The results
of our evaluations (Table 2)  confirm the existence of  $P-$wave 
\textbf{\ }$\overline{q}q^{\prime }-$mesons with masses $\sim 980$ $MeV$.
Another fact in favour  of
 $P-$wave \textbf{\ }$\overline{q}q^{\prime }-$mesons with masses $\sim
980$ $MeV$ is the coincidence of the evaluated mass  for the first
radial excitation of $a_{0}(980)$ meson  with the mass value of the 
$a_{0}(1450)$ meson \cite{eidel}. This meson has been predicted 
for the first time  in the framework of finite energy
QCD sum rules  \cite{gor}.



\hspace*{-0.5cm}
{\small Table 1. }

\hspace*{-0.5cm}
{\small\parbox[t]{13.9cm}{Evaluated masses  of the $1^{--}$ and $0^{-+}$ 
mesons in comparison with the  data from Ref. \cite{eidel}.}}
\begin{center}
\begin{tabular}{|c|c|c|c|c|c|}
\hline
Meson & $M_{n}^{exp}[MeV]$ & $M_{n}^{th}[MeV]$ & 
Meson & $M_{n}^{exp}[MeV]$ & $M_{n}^{th}[MeV]$ \\ 
\hline
$\rho $ & 775.8$\pm $0.5 & 740 & $B$ & 5279.2$\pm $0.5 & 5250  \\ 
$\rho ^{\prime }$ & 1465$\pm $25 & 1455 & $B_{s}$ & 5369.6$\pm $2.4 & 5370\\ 
$\rho ^{\prime \prime }$ & 1720$\pm $20 & 1730 & $B_{c}$ & 6400$\pm $400 & 6450  \\ 
$\phi $ & 1019.456$\pm $0.020 & 1010 & $\eta _{b}$ & - & 9330 \\ 
$\phi ^{\prime }$ & 1680$\pm $20 & 1650 &  $\pi ^{\prime }$ & 1300$\pm $100 & 1290\\ 
$\phi ^{\prime \prime }$ & - & 2050 & $K^{\prime }$ & - & 1400\\ 
$J/\psi $ & 3096.916$\pm $0.011 & 3060 & $D^{\prime }$ & - & 2450\\ 
$\psi ^{\prime }$ & 3686.093$\pm $0.034 & 3650 & $D_{s}^{\prime }$ & - & 2560\\ 
$\psi ^{\prime \prime }$ & 4040$\pm $10 & 4070 & $\eta _{c}^{\prime }$ 
& - & 3600\\ 
$\psi ^{\prime \prime \prime }$ & 4415$\pm $6 & 4390 &$B^{\prime }$ & - & 5650 \\ 
$\Upsilon $ & 9460.30$\pm $0.26 & 9470 & $B_{s}^{\prime }$ & - & 5750\\ 
$\Upsilon ^{\prime }$ & 10023.26$\pm $0.31 & 9990 &  $B_{c}^{\prime }$ & - & 6800 \\ 
$\Upsilon ^{\prime \prime }$ & 10355.2$\pm $0.5 & 10325 & $\eta _{b}^{\prime }$ & - & 9960 \\ 
$\Upsilon ^{\prime \prime \prime }$ & 10580$\pm $3.5 & 10550 & $\pi
^{\prime \prime }$ & 1812$\pm $14 & 1810 \\ 
$\Upsilon ^{5S}$ & 10865$\pm $8 & 10830 & $K^{\prime \prime }$ & -
& 1850 \\ 
$\Upsilon ^{6S}$ & 11019$\pm $8 & 10985 & $D^{\prime \prime
} $ & - & 2880 \\ 
$\pi $ & 138.039$\pm $0.004 & 120 & $%
D_{s}^{\prime \prime }$ & - & 2950 \\ 
$K$ & 495.66$\pm $0.02 & 500 & $\eta _{c}^{\prime \prime }$ & - & 3970 \\ 
$D$ & 1867.0$\pm $0.5 & 1850& $B^{\prime \prime }$
& - & 6070 \\ 
$D_{s}$ & 1968.3$\pm $0.5 & 1990& $%
B_{s}^{\prime \prime }$ & - & 6130 \\ 
$\eta _{c}$ & 2979.6$\pm $1.2 & 2990& $%
B_{c}^{\prime \prime }$ & - & 7150  \\
\hline
\end{tabular}

\end{center}

\medskip

\hspace*{-0.5cm}
{\small Table 2. }

\hspace*{-0.5cm}
{\small 
\parbox[t]{15.cm}{Evaluated masses in MeV for the ground states  and the
orbital excitations of the   $\bar qq'-$ mesons in
comparison with the  data from Ref.\cite{eidel}. }}
\begin{center}
\begin{tabular}{cccccccc}
\hline
Meson & $J^{PC}$ & $M^{exp}(MeV)$ & $M^{ev}(MeV)$ & Meson & $J^{PC}$
& $M^{exp}(MeV)$ & $M^{ev}(MeV)$ \\ \hline
$\pi $ & 0$^{-+}$ & 138.039$\pm $0.004 & 138$\pm $30 & $\rho _{3}$ & 3$^{--}$ & 
1688.8$\pm $2.1 & 1722$\pm $30 \\ 
$\rho $ & 1$^{--}$ & 775.8$\pm $0.5 & 770$\pm $30 & $a_{2}^{3}$ & 2$^{++}$ & 
- & 1873$\pm $30 \\ 
$a_{0}$ & 0$^{++}$ & 984.7$\pm $1.2 & 998$\pm $30 & $b_{3}^{3}$ & 3$^{+-}$ & 
- & 2024$\pm $30 \\ 
$b_{1}$ & 1$^{+-}$ & 1229.5$\pm $3.2 & 1227$\pm $30 & $a_{3}^{3}$ & 3$^{++}$
& - & 2031$\pm $30 \\ 
$a_{1}$ & 1$^{++}$ & 1230$\pm $40 & 1280$\pm $30 & $a_{4}$ & 4$^{++}$ & 2010$%
\pm $12 & 2037$\pm $30 \\ 
$a_{2}$ & 2$^{++}$ & 1318.3$\pm $0.6 & 1334$\pm $30 & $a_{3}^{4}$ & 3$^{--}$
& - & 2177$\pm $30 \\ 
$a_{1}^{2}$ & 1$^{--}$ & - & 1506$\pm $30 & $b_{4}^{4}$ & 4$^{-+}$ & - & 2316%
$\pm $30 \\ 
$\pi _{2}$ & 2$^{-+}$ & 1672.4$\pm $3.2 & 1678$\pm $30 & $a_{4}^{4}$ & 4$%
^{--}$ & - & 2313$\pm $30 \\ 
$a_{2}^{2}$ & 2$^{--}$ & - & 1700$\pm $30 & $a_{5}^{4}$ & 5$^{--}$ & - & 2310%
$\pm $30 \\ \hline
\end{tabular}
\end{center}

\end{document}